\documentclass[3p,times]{elsarticle}

\usepackage{ecrc}

\usepackage{color}
\volume{00}

\firstpage{1}

\journalname{Nuclear Physics A}

\runauth{D.~Cabrera et al.}

\jid{NUPHA}

\jnltitlelogo{Nuclear Physics A}

\CopyrightLine{2012}{Published by Elsevier Ltd.}

\usepackage{amssymb}

\usepackage[figuresright]{rotating}

\newcommand{\be}{\begin{equation}} 
\newcommand{\ee}{\end{equation}} 
\newcommand{\ba}{\begin{eqnarray}} 
\newcommand{\ea}{\end{eqnarray}}

\begin{document}

\begin{frontmatter}

\dochead{}



\title{Heavy flavor relaxation in a hadronic medium}

\author[a,b,c]{Daniel Cabrera}\author[d]{Luciano M.~Abreu}\author[e]{Felipe
J.~Llanes-Estrada}\author[e,f]{\\ Juan M.~Torres-Rincon}

\address[a]{Departamento de F\'{\i}sica Te\'orica II, Universidad Complutense, 28040
Madrid, Spain}

\address[b]{Institute for Theoretical Physics, Frankfurt University, 60438
Frankfurt-am-Main, Germany}

\address[c]{Frankfurt Institute for Advanced Studies, Frankfurt University,
60438 Frankfurt-am-Main, Germany}

\address[d]{Instituto de F\'{\i}sica, Universidade Federal da Bahia, 40210-340,
Salvador, BA, Brazil}

\address[e]{Departamento de F\'{\i}sica Te\'orica I, Universidad Complutense, 28040
Madrid, Spain}

\address[f]{Instituto de Ciencias del Espacio (IEEC/CSIC), Campus Universitat
Aut\`onoma de Barcelona, Facultat de Ci\`encies, Torre C5, E-08193 Belalterra
(Barcelona), Spain}

\begin{abstract}
Charm and bottom transport coefficients in a medium constituted of light mesons, such as
is formed in the hadronic phase of Heavy Ion Collisions, are obtained within an
effective field theory approach implementing heavy quark symmetry and
chiral symmetry breaking.
Heavy flavor propagates in the medium as $D/B$ and $D^*/B^*$ degrees of freedom, and unitarization of the lowest order heavy-light meson amplitudes is used in order to
reach high temperatures. The latter accounts for dynamically generated resonances in isospin $1/2$ channels, a feature that leads to a more efficient heavy flavor diffussion.
We discuss the temperature and momentum dependence of the friction and diffusion coefficients in
a transport approach up to temperatures of about $T\simeq 150$ MeV,
and provide
estimates of the charm/bottom relaxation lengths and momentum loss.
Implications for heavy meson spectrum observables in Heavy Ion
Collisions are discussed.
\end{abstract}


\end{frontmatter}


\section{Introduction}
\label{sec:intro}

The features of matter formed in Heavy Ion Collisions (HICs) have been a subject
of great interest in the last decades. In this scenario, heavy--flavored hadrons
play an essential role since they carry heavy quarks produced in the early stage
of the collisions, unlike pions and kaons which
can be produced in the thermal medium at later stages. 

It is worth noticing that the momentum spectra of charmed and bottomed
mesons extracted from HICs undergo modifications due to their interactions with
the hadronic medium. In this sense, the
diffusion of heavy mesons in an equilibrium hadronic gas should be taken into account in order to compute realistic spectra, for instance, in transport approaches. Unlike other lighter systems, relatively little attention has been devoted to study heavy-meson dynamics in the hadronic phase of a HIC. In addition, there is a considerable dispersion in existing results on the matter, as for instance concerning the value of the charm relaxation length at a given temperature \cite{Laine:2011is,He:2011yi,Ghosh:2011bw,LADCFLEJT:2011,He:2011qa,Das:2011}.

In this talk we report on recent progress in the calculation of the transport coefficients of charmed and bottomed mesons in a hot gas consisting of pions, kaons and $\eta$ mesons. In order to do this we provide a realistic determination of the relevant heavy-light meson scattering amplitudes in an effective field theory approach, exploiting chiral perturbation theory at next-to-leading order
(NLO) together with constraints from heavy quark symmetry, and implementing exact unitarization in order to reach higher
temperatures. The latter is a crucial point in our analysis, leading to dynamically generated states in agreement with our current knowledge of the $D$ and $B$ spectra with a minimal number of parameters, and therefore to a more realistic (and theoretically consistent) determination of the transport coefficients.

\section{Modelling heavy-meson interactions in a light meson gas}
\label{sec:interaction}


We calculate heavy-meson transport coefficients in a Fokker-Plank equation approach. The dynamical input is encoded in the scattering amplitudes of (long-lived) charmed and bottomed mesons with the octet of light pseudoscalar mesons. We assume that
the density of heavy mesons is negligible and thus ignore 
collisions among themselves. For a detailed discussion of the model for charm interactions within the pion gas we refer to our previous work~\cite{LADCFLEJT:2011}.

The amplitude for scattering off a charm (bottom) quark in the light meson
gas, at next-to-leading order (NLO) in the chiral expansion and leading order (LO) in the heavy quark expansion,
irrespective of whether the heavy quark is in a $D(B)$ or a $D(B)^*$ meson state, is given
by
\ba
\label{ampl} {\cal V} &=&   \frac{C_{0}}{4 F^2} (s - u)  + \frac{2C_{1}}{ F^2} h_1    +  \frac{ 2 C_{2}}{ F^2} h_3 (p_2 \cdot p_4 ) 
+  \frac{2 C_{3}}{ F^2} h_5 \left[ (p_1 \cdot p_2 ) (p_3 \cdot p_4 ) + (p_1 \cdot
  p_4 )(p_2 \cdot p_3 ) \right] \ ,
\ea
where $ C_{i}\;(i=0,...,3)$ are channel-dependent numerical coefficients in
isospin basis. The three (flavor-dependent) parameters $h_i$ are the low-energy constants (LECs) from the NLO chiral Lagrangian, which we constrain with the available experimental information on the heavy-meson spectrum (a list of the relevant states is accounted for in Table~\ref{tab:spectrum}).


The chiral perturbative expansion is typically bound to work properly only at very low energies, and cannot describe the appearance of resonances in a given scattering channel. In addition, at high energies, perturbative cross sections violate Froissart bounds imposed by
the unitarity of the $S$-matrix.
We solve these limitations by implementing exact unitarity in the scattering amplitudes by solving the Bethe-Salpeter equation with the perturbative amplitudes as dynamical kernel, namely
\be 
T (s) = -V (s)/(1- V (s) \ G (s))\ , 
\label{Unitarizedampl}
\ee
where $V (s)$ is the $S$-wave projection of the scattering amplitude in Eq.
(\ref{ampl}), and $G (s)$ stands for the two-meson resolvent function (loop integral), conveniently regularized.


The free parameters of the model include a subtraction constant from dimensional regularization of $G(s)$, plus the three LECs. The subtraction constant is fixed by the resonance position in the isospin $1/2$ channels. Note that these resonances are already generated by the LO in the interaction [$(s-u)$ term in Eq.~(\ref{ampl})], which is completely determined by chiral symmetry breaking, and the freedom provided by the LECs can be used to improve on the LO results. They have to be determined for the charm and bottom sector, although some of the LECs scale with the heavy meson mass and thus not all of them are independent. Note that the bottom sector is much less known experimentally: Particularly, the $S$-wave resonances have not yet been observed. Given the fact that the two $D_1$ states are very close in energy, we assume that the same holds for the bottom sector, provided that heavy quark symmetry works. Also note that the $D(B)\pi$ and $D(B)^*\pi$ amplitud
 es are related in the heavy quark limit just by exchanging $m_{D(B)}$ by $m_{D(B)^*}$ in Eq.~(\ref{Unitarizedampl}). For the bottom sector, we obtain as our best determination $m_{B_0}=5534$~MeV ($\Gamma_{B_0}=210$~MeV) and $m_{B_1}=5587$~MeV ($\Gamma_{B_1}=250$~MeV) in good agreement with \cite{Guo:2006fu,Guo:2006rp,Kolomeitsev:2003ac} (c.f.~\cite{LADCFLEJT:2011} for the charm sector). Threshold cross sections (or scattering lengths) are also obtained in good agreement with previous determinations.

\begin{table}[htbp]
\begin{center}
\begin{tabular}{|c ||c |c|c|c||c|c|c|c|}
\hline 
Spin,  $J^\pi$  & D (QM) & D (exp.) & (M, $\Gamma$) MeV & this work & B (QM) & B (exp.) &(M, $\Gamma$) MeV & this work\\
\hline
1/2, $0^+$ & $D_0$ & $D_0^* (2400)$ & 2318, 267 & 2300, 350 & $B_0$ & ? & ? & 5534, 210 \\
1/2, $1^+$ & $D_1$ & $D_1 (2430)$ & 2427, 384 & 2390, 400 & $B_1$ & ? & ? & 5587, 250 \\
3/2, $1^+$ & $D_1$ & $D_1(2420)^0$ & 2421, 27 & ... & $B_1$ & $B_1$(5721) &  5723, ? & ...\\
3/2, $2^+$ & $D_2$ & $D^*_2 (2460)$ & 2466, 49 & ... & $B_2$ & $B^*_2$ (5747) & 5743, 23 & ... \\
\hline
\end{tabular}
\caption{ \label{tab:spectrum} List of heavy meson states. Left side: $D$ meson. Right side: $B$ mesons. The spin $1/2$ doublets are the ones dynamically generated from unitarization of the $S$-wave interaction. Experimental
data from \cite{Beringer:1900zz}. Lack of experimental evidence is labeled with a question mark.}
\end{center}
\end{table}

\section{Heavy-meson transport coefficients}
\label{sec:transport}

The momentum-space distribution of bottomed mesons must relax via the Fokker-Planck equation (we refer to \cite{LADCFLEJT:2011} for details and assumptions). 
Let us label the momenta of the elastic collision between a heavy meson $M^{(*)}$ and a
light meson $\phi$ as
$M ^{(*)}(\mathbf{p}) + \phi (\mathbf{q}) \rightarrow
M^{(*)}(\mathbf{p}-\mathbf{k}) + \phi (\mathbf{q} +\mathbf{k})$.
The evolution of the momentum distribution of the heavy meson
due to its
interaction with the isotropic mesonic gas is controlled by the drag ($F$) and
diffusion ($\Gamma _0,\Gamma_1$) coefficients, written as 
\begin{eqnarray} \label{Transportintegrals}
 F(p^2)  =  \int d\mathbf{k}\  w(\mathbf{p},\mathbf{k})  \ \frac{k_ip^i}{p^2} \ , \quad
 \Gamma_0 (p^2)  =   \frac{1}{4} \int d\mathbf{k}\ w(\mathbf{p},\mathbf{k}) \left[ \mathbf{k}^2 - \frac{(k_i p^i)^2}{p^2} \right] \ , \quad
 \Gamma_1(p^2)  =   \frac{1}{2} \int d\mathbf{k}\  w(\mathbf{p},\mathbf{k}) \
 \frac{(k_i p^i)^2}{p^2} \ ,
\end{eqnarray}
where $w (\mathbf{p}, \mathbf{k} )$ is the collision rate for a heavy meson
with initial (final) momentum $\mathbf{p}$ ($\mathbf{p}-\mathbf{k}$), 
\be 
w (\mathbf{p}, \mathbf{k} ) = g_{\phi} \int \frac{d \mathbf{q}}{(2\pi)^9} f_{\phi} (\mathbf{q}) \left[ 1+ f_{\phi} (\mathbf{q}+\mathbf{k}) \right]
  \frac{1}{2E_p^M}  \frac{1}{2E_q^{\phi}} \frac{1}{2 E_{p-k}^M} \frac{1}{2 E_{q+k}^{\phi}}
 (2\pi)^4 \delta (E_p^M + E_q^{\phi} - E^M_{p-k} -E_{q+k}^{\phi} )  \sum |\mathcal{M}_{M \phi }(s,t,\chi)|^2 . 
 \label{probdist}
\ee
\noindent
Here $f_{\phi} (\mathbf{q})$ is the bath's distribution
function, $\mathcal{M}_{M \phi }$ stands for the Lorentz invariant heavy-light meson 
scattering matrix element, $g_\phi$ is the Goldstone boson isospin
degeneracy (e.g. $g_\pi=3$ for the pion), and $\chi$ denotes possible spin
degrees of freedom. Since spin average is trivial in view of heavy quark symmetry, one just has
$
\sum | \mathcal{M}_{M \phi } (s,t,\chi)|^2 = \frac{1}{\sum_{I} (2I+1)}\sum_{I}  (2I+1) |T^{I}|^2
$, in terms of the isospin averaged unitarized amplitude.

\begin{figure}
\centering
\includegraphics[width=0.32\textwidth]{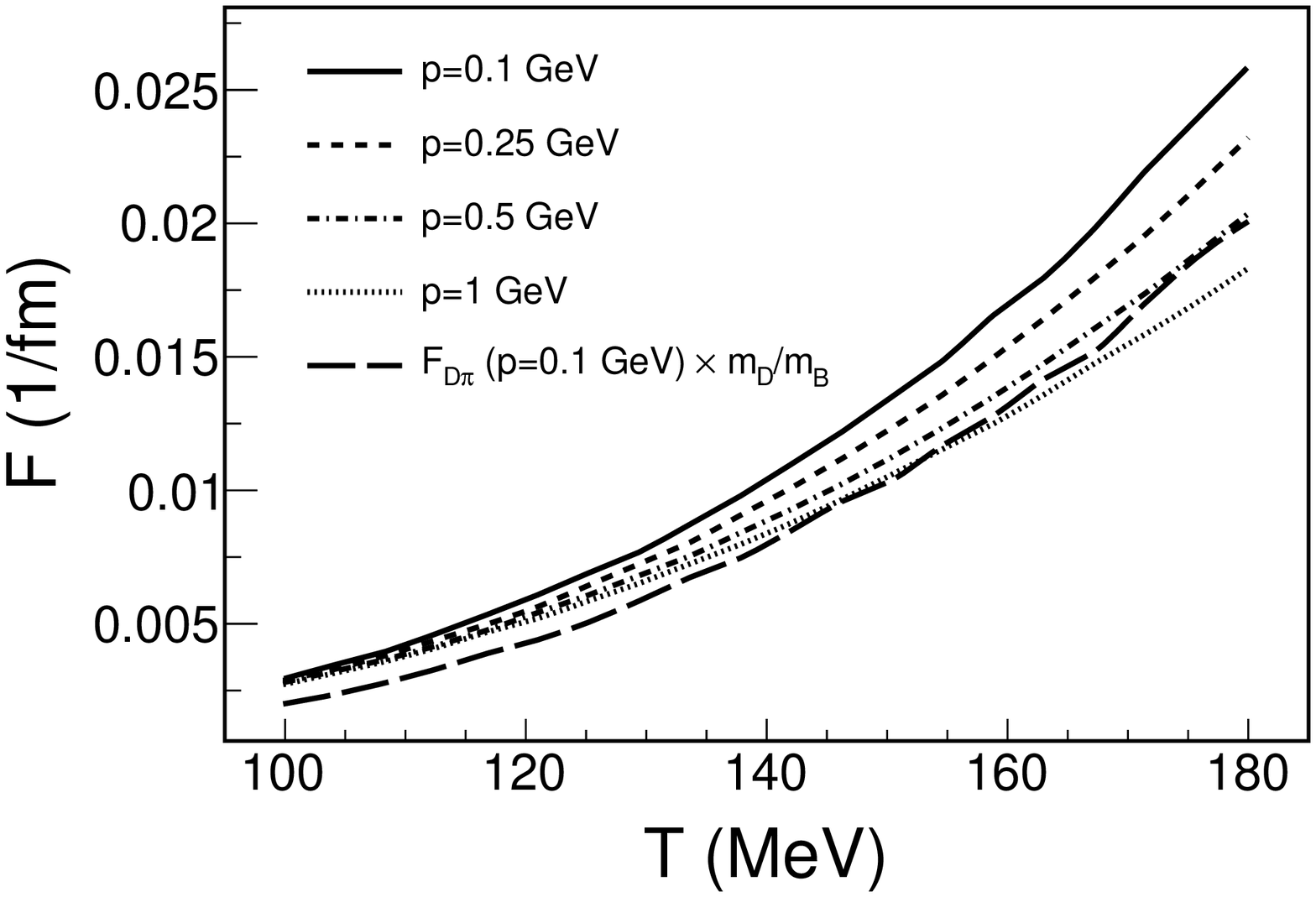}
\hspace{0.1cm}
\includegraphics[width=0.32\textwidth]{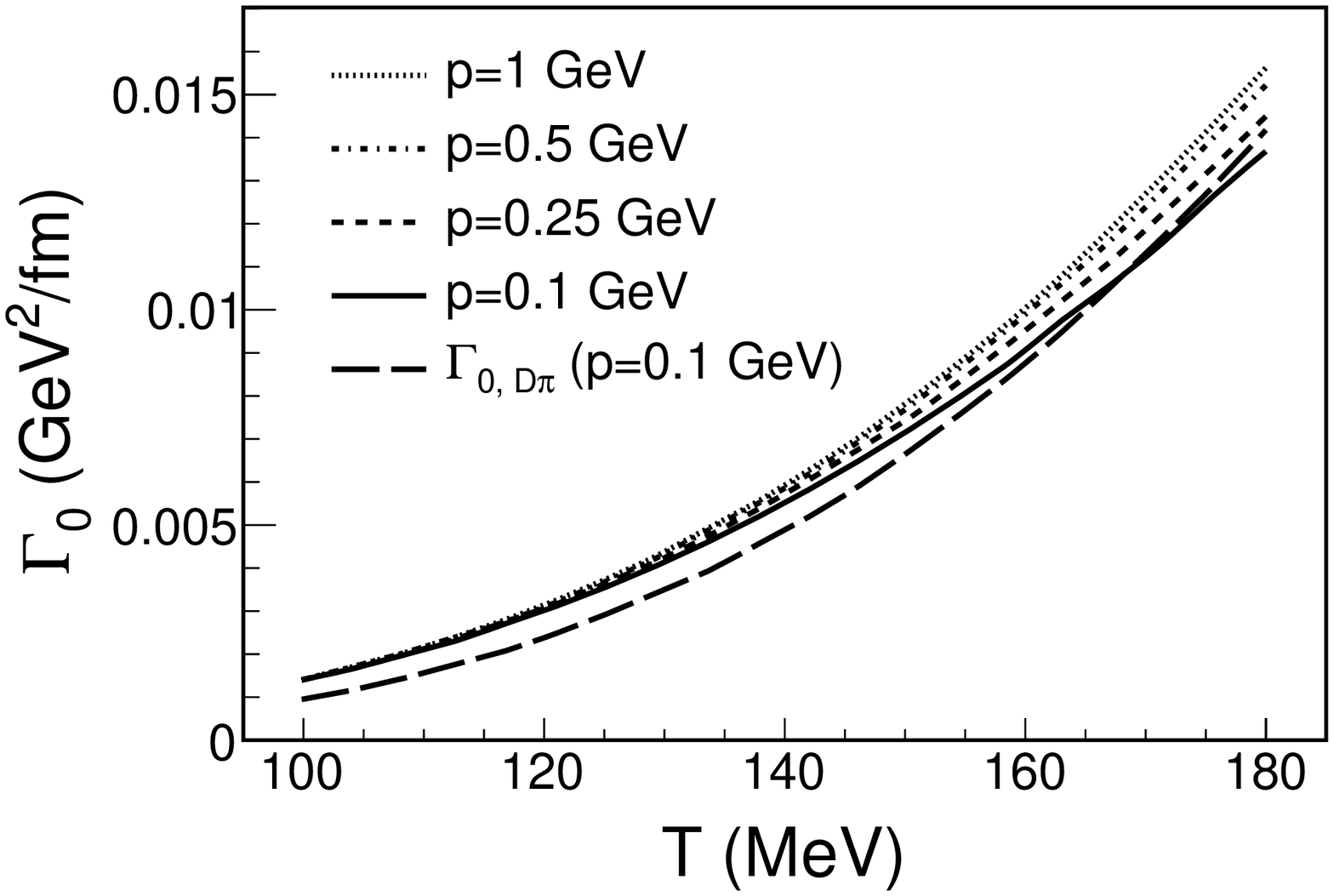}
\hspace{0.1cm}
\includegraphics[width=0.32\textwidth]{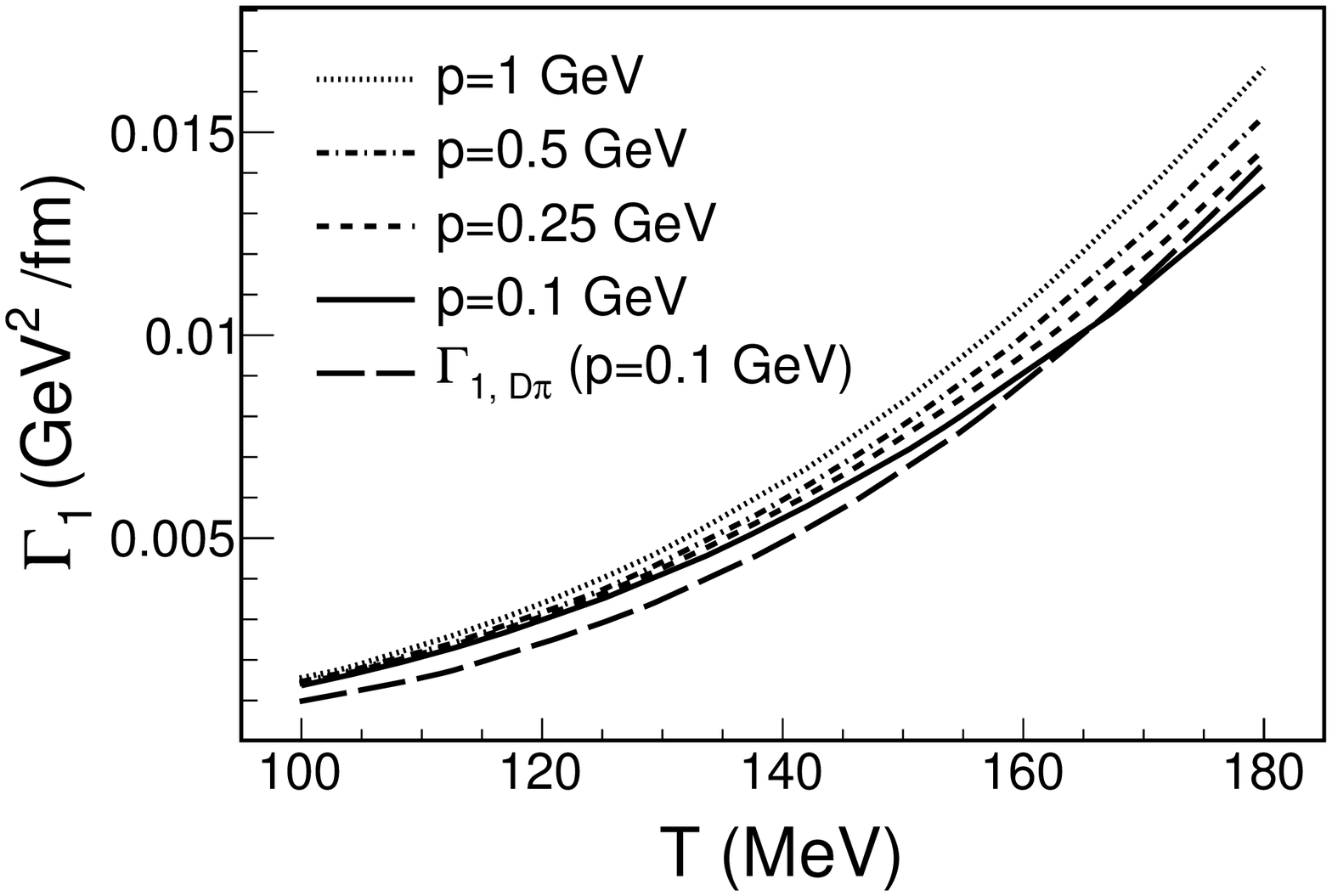}
\caption{\label{fig:FG0G1-pion-gas-vs-T} 
Momentum-space drag ($F$) and diffusion ($\Gamma_0$, $\Gamma_1$) coefficients as a function of temperature for several bottom quark momenta in the pion gas. The charm coefficients are shown for comparison at $p=0.1$~GeV \cite{LADCFLEJT:2011}.}
\end{figure}

In Fig.~\ref{fig:FG0G1-pion-gas-vs-T} we show the temperature and momentum dependence of the $B$-meson
drag and diffusion coefficients. We observe an increase of factor 6-8 in the range $T=100-180$~MeV, namely the drag in a heavy-ion collision is considerably
strengthened in the hotter stages, with significant interaction between heavy
mesons and the thermal medium (and accordingly at larger momentum transfers).
From $F$ we can estimate the relaxation length of bottom quarks in the
hadronic medium. In the pion gas at $T=150$~MeV we find
$
\lambda_B (T= 150 \textrm{ MeV}, p = 1 \textrm{ GeV}) = \frac{1}{F}\simeq  100\textrm{ fm}
$
for a typical momentum of 1~GeV, to be compared to $\lambda_D\simeq 40$~fm for charmed mesons evaluated in the same approach \cite{LADCFLEJT:2011}.
Therefore bottomed mesons barely relax during the lifetime of the 
hadron gas, unlike charm mesons that, while not relaxing completely, may loose a great deal of memory of the initial state. In this sense, bottomed mesons constitute an optimal system to characterize the early stages of a relativistic heavy-ion
collision, in spite of their (detected) number being less abundant than for charm.
Also note that some of the observed $c$ quarks in the final state proceed from $b$ quark weak decays in the micrometer range, and therefore they interacted as $b$ quarks with the (hadronic) medium.

We observe that, in the static limit ($p\to 0$), the drag coefficient for bottomed mesons is about 3 times smaller than for the charm case, whereas the diffusion coefficients are very similar for the two systems, c.f.~Fig.~\ref{fig:FG0G1-pion-gas-vs-T}. This scaling is the correct one as expected from non-relativistic kinetic theory, $F \simeq \frac{1}{3} \sigma P \sqrt{\frac{m_{\pi}}{T}} \frac{1}{m_B}$ and $\Gamma_0 , \Gamma_1 \simeq \frac{1}{3} \sigma P \sqrt{m_{\pi} T}$ ($\sigma$ is the total cross section and $P$ is the pressure of the gas). Such result constitutes a consistency test of our calculation and is also observed for the transport coefficients of heavy quarks beyond the critical temperature \cite{Rapp:2008qc,vanHees:2007me}.
We have found that implementing unitarity to obtain realistic amplitudes for heavy meson scattering off the light gas plays an important role and has a strong impact in the final size and temperature dependence of the transport coefficients, as well as on their scaling properties with the heavy quark mass.
For instance, the use of perturbative cross sections overestimates scattering at high energies and leads to an unrealistically large drag force \cite{Ghosh:2011bw}. Alternatively, scattering lengths have also been used as dynamical input \cite{Das:2011}, which underestimates difussion due to the $s$-channel enhancement of the interaction in the resonant channels. The unitarized scheme provides the most realistic result, coherently with our current knowledge of the heavy meson spectrum and with a controlled high energy behavior.

\begin{figure}
\centering
\includegraphics[width=0.33\textwidth]{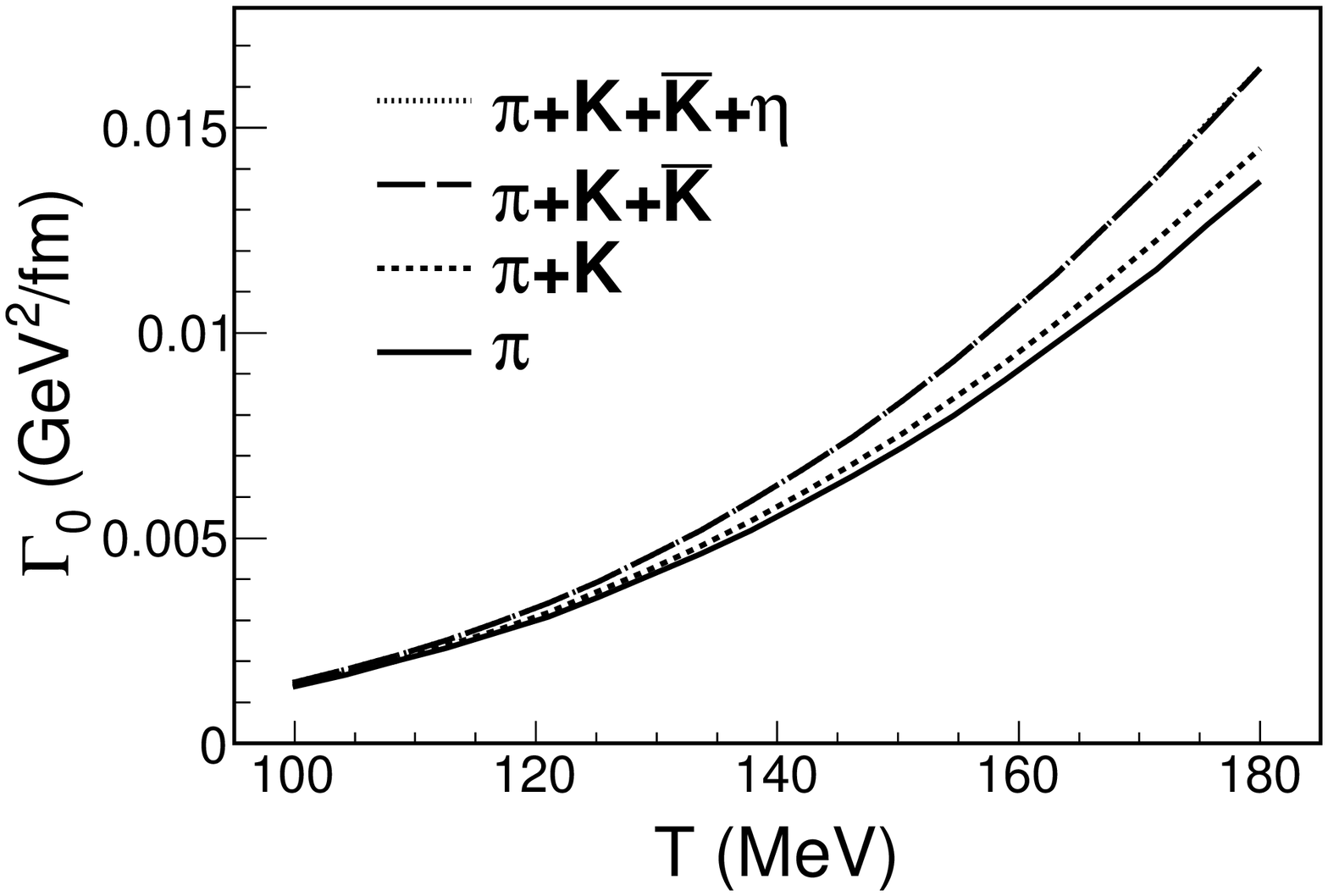}
\includegraphics[width=0.33\textwidth]{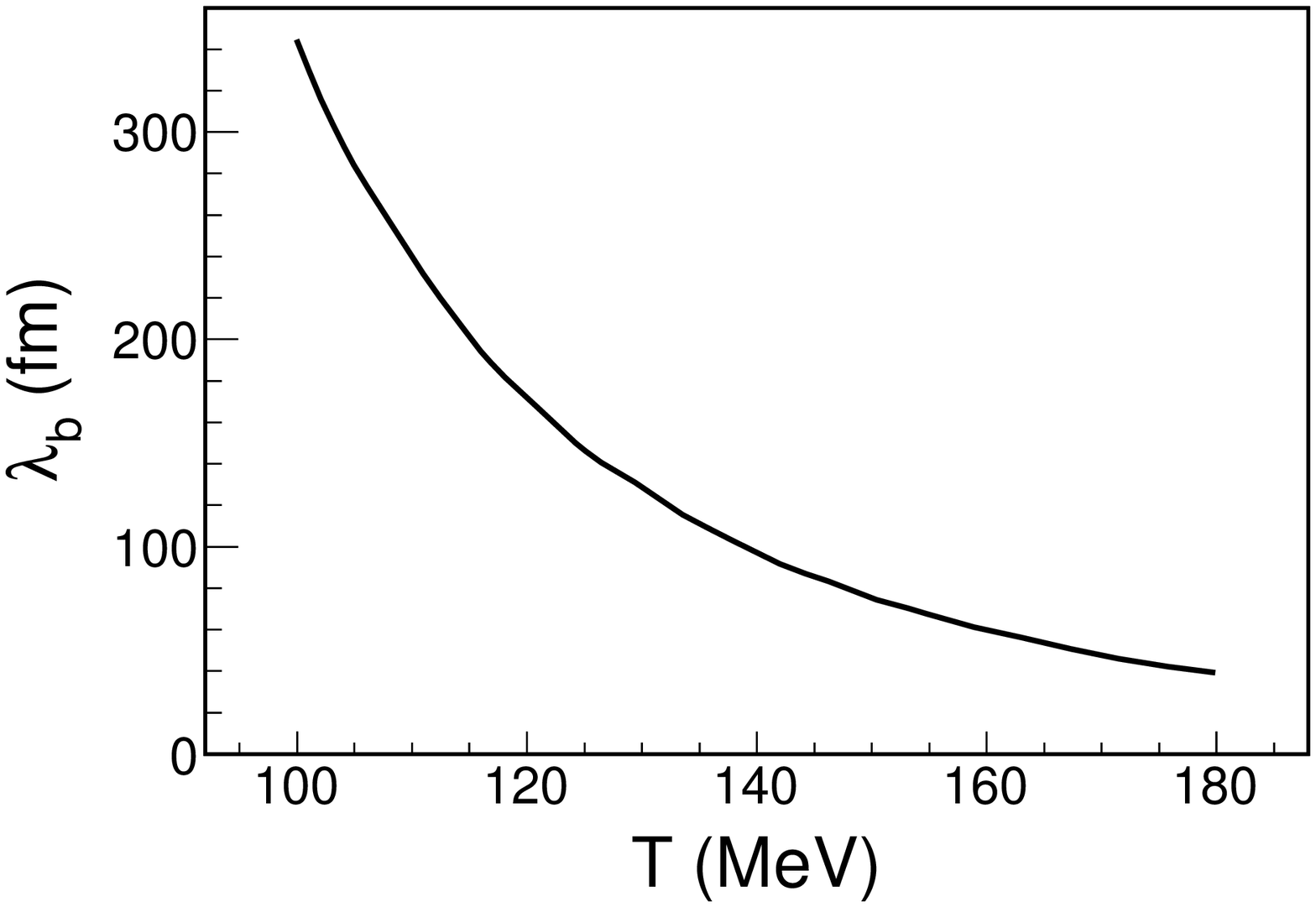}
\includegraphics[width=0.33\textwidth]{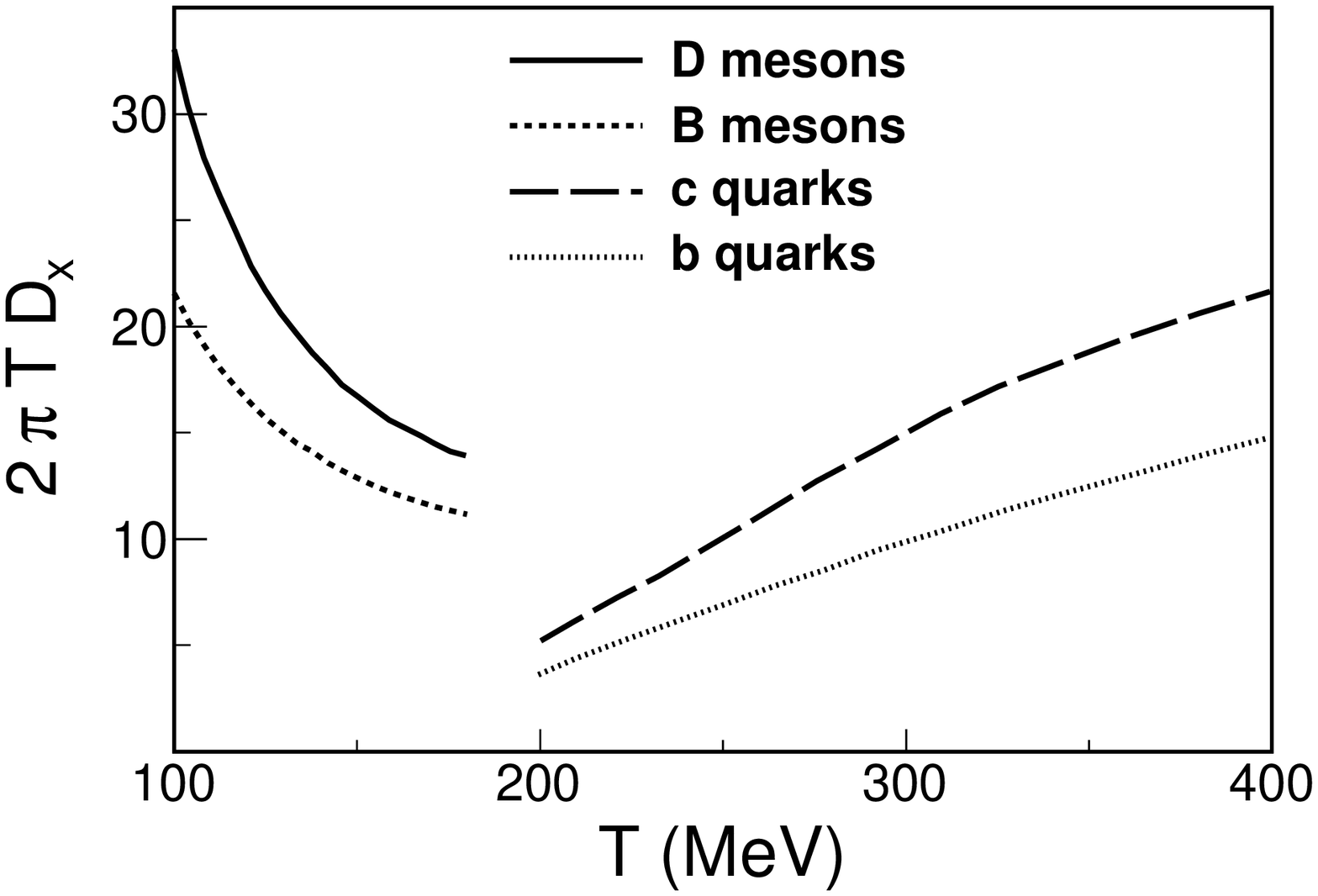}
\caption{\label{fig:lambda-G0-Dx-full-gas} 
Left: Contributions to the $\Gamma_0$ coefficient in the $SU(3)$ meson gas at $p=0.1$ GeV. Center: Bottom relaxation length in the hadronic phase at $p=0.1$~GeV as a function of the temperature. Right: Charm and bottom spacial diffusion coefficients below (this work) and above (\cite{Rapp:2008qc,vanHees:2007me,He:2012df}) the crossover.}
\end{figure}

We analyze the contributions from the different species in the $SU(3)$ gas for a bottomed quark in Fig.~\ref{fig:lambda-G0-Dx-full-gas}~(left).
As expected, the most relevant contribution comes from the pion gas. At $T=150$~MeV, 
pions provide almost 90\% of the total, while the next-to-leading contribution is
provided by kaons and (mostly) anti-kaons.
The relaxation length of
bottomed mesons travelling with 1~GeV momentum in the full meson gas is
$
\lambda_B (T= 150 \textrm{ MeV}, p = 1 \textrm{ GeV}) = \frac{1}{F}\simeq  80\textrm{ fm}
$, c.f.~Fig.~\ref{fig:lambda-G0-Dx-full-gas}~(center).
Still,
this thermal relaxation time is greater than the lifetime of the hadron gas,
reinforcing the idea that heavy quarks are optimal carriers of information of
the phase transition upon exiting the interacting region.
Our estimate of the spacial diffusion coefficient for charmed and bottomed mesons, Fig.~\ref{fig:lambda-G0-Dx-full-gas}~(right), together with estimates in the quark-gluon plasma phase, indicates that the relaxation time of heavy quarks has a minimum around the crossover.

Finally, a momentum loss per unit length of about 50~MeV/fm (70~MeV/fm) is obtained in our approach for a charmed (bottomed) meson propagating in the gas with a typical momentum of 1~GeV with respect to the rest frame of the medium.
The results discussed above allow us to estimate the modification in heavy quark spectra due to diffusion in the hadronic phase.
The observed distribution of $D$ mesons, being different from the one after hadronization, may lead to a different determination of the freeze-out temperature from fits to the Botzmann exponential shape at low $p_T$.
Observables like the nuclear suppression factor or the elliptic flow of heavy flavored mesons can be sensitive, in particular, to the sizable momentum dependence of the transport coefficients found in this work, which could be implemented in hydrodynamical simulations \cite{Lang:2012vv}.

\section{Acknowledgments}
We want to thank Feng-Kun Guo, Juan Nieves, Santosh Ghosh, Christine Davies and Rachel Dowdall for clarifications
and comments.
We acknowledge financial support from grants
FPA2011-27853-C02-01, FPA2011-27853-C02-02,FIS2008-01323 (Ministerio de Econom\'{\i}a y Competitividad, Spain) and from the EU Integrated
Infrastructure Initiative Hadron Physics Project under Grant Agreement
n.~227431. LMA thanks CAPES (Brazil) for partial finantial support. 
DC acknowledges financial support from Centro Nacional de
F\'isica de Part\'iculas, Astropart\'iculas y Nuclear
(CPAN, Consolider-Ingenio 2010) postdoctoral programme.
JMTR is a recipient of an FPU grant (Ministerio de Educaci\'on, Cultura y Deporte, Spain).








\end{document}